\documentclass[pra ,aps, reprint, superscriptaddress, a4paper]{revtex4-1}
\usepackage{physics}
\usepackage{amsmath,amssymb}
\usepackage{braket}
\usepackage{graphicx}
\usepackage{xcolor}
\usepackage{verbatim}
\usepackage{color}
\usepackage[colorlinks, linkcolor={blue}, citecolor={blue}, urlcolor={blue}]{hyperref}
\usepackage{tabularx}
\usepackage{booktabs}
\usepackage{layouts}
\usepackage{blindtext}
\newcommand{\munich}{Department of Physics and Arnold Sommerfeld Center for Theoretical Physics (ASC), Ludwig-Maximilians-Universit\"at M\"unchen, Theresienstr.~37,  D-80333 M\"unchen, Germany} 
\newcommand{\mcqst}{Munich Center for Quantum Science and Technology (MCQST), Schellingstr.~4, D-80799 M\"unchen, Germany}
\begin{document}
\title{
Snapshot based characterization of particle currents\\ and the Hall response in synthetic flux lattices
}
\author{Maximilian Buser}
\affiliation{\munich}
\affiliation{\mcqst}
\author{Ulrich Schollw\"ock}
\affiliation{\munich}
\affiliation{\mcqst}
\author{Fabian Grusdt}
\affiliation{\munich}
\affiliation{\mcqst}
\date{\today}
\begin{abstract}
Quantum simulators are attracting great interest because they promise insight into the behavior of quantum many-body systems that are prohibitive for classical simulations.
The generic output of quantum simulators are snapshots, obtained by means of projective measurements.
A central goal of theoretical efforts must be to predict the exact same quantities that can be measured in experiments.
Here, we report on the snapshot based calculation of particle currents in quantum lattice models with a conserved number of particles.
It is shown how the full probability distribution of locally resolved particle currents can be obtained from suitable snapshot data.
Moreover, we investigate the Hall response of interacting bosonic flux ladders, exploiting snapshots drawn from matrix-product states.
Flux ladders are minimal lattice models, which enable microscopic studies of the Hall response in correlated quantum phases and they are successfully realized in current quantum-gas experiments.
Using a specific pattern of unitary two-site transformations, it is shown that the Hall polarization and the Hall voltage can be faithfully computed from the snapshots obtained in experimentally feasible quench and finite-bias simulations.
\end{abstract}	
\maketitle
\section{Introduction}
Quantum simulators are attracting great interest because they promise insight into the behavior of quantum many-body systems that are prohibitive for classical simulations~\cite{georgescu_14}.
In addition to trapped ions~\cite{blatt_12} and superconducting qubits~\cite{wilkinson_20}, ultracold atoms in optical lattices are a particularly successful platform for this purpose~\cite{bloch_08, bloch_12, gross_17}.
The generic output of any quantum simulator are snapshots, which are taken in the framework of projective measurements.
For instance, quantum-gas microscopes for ultracold atoms in optical lattices offer single-atom, single-site, and spin-state resolution~\cite{bakr_09, sherson_10, haller_15, hartke_20, koepsell_20}. 
They enable measurements of non-local observables, which are inaccessible in different physical platforms~\cite{hilker_17, mazurenko_17, koepsell_19, koepsell_20b}.
%

%
In view of the experimental advances, the benchmarking and verification of quantum simulators by means of powerful classical algorithms has become a difficult and important task~\cite{eisert_20}.
Predicting the snapshot characteristics that can be measured in experiments must be a central goal of theoretical efforts.
In recent years, data science tools and machine learning techniques have proven effective for the analysis of snapshots obtained from real quantum-gas simulators~\cite{chiu_19, rem_19, bohrdt_19, miles_20, bohrdt_20, bohrdt_21}.
Moreover, theoretical snapshots offer great flexibility and they are a key resource for the further development of machine learning approaches in the context of quantum lattice models~\cite{ferris_12, torlai_18, carrasquilla_19, humeniuk_21, golubeva_21, vieijra_21}.
In this article, we demonstrate how state-of-the-art numerical matrix-product-state algorithms can be used to generate snapshot data to model realistic experimental settings of current interest.
%

%
Matrix-product states are particularly efficient representations of quantum-lattice wave functions, especially for the ground states of one-dimensional systems.
They are at the heart of successful algorithms for the classical simulation of quantum systems, such as the density-matrix renormalization-group method~\cite{white_92, schollwoeck_05, schollwoeck_11} and state-of-the-art time-evolution algorithms~\cite{haegeman_11, paeckel_19}.
Moreover, it has been shown that snapshots can be efficiently sampled from matrix-product states~\cite{ferris_12}.
Overall, theoretical snapshots obtained from matrix-product states enable the benchmarking and verification of real quantum devices and they allow to better evaluate the feasibility of theoretical proposals.
%

%
As a concrete physical system, we will consider flux ladders.
These are minimal lattice models which allow to study the rich interplay between effective magnetic fields and interactions among quantum particles. 
Because of this interplay, flux-ladder models host a myriad of ground-state phases~\cite{orignac_01, carr_06, roux_07, dhar_12, petrescu_13, huegel_14, tokuno_14, uchino_15, piraud_15, barbarino_15, greschner_15, didio_15, petrescu_15, kolley_15, cornfeld_15, ghosh_16, greschner_16, orignac_17, strinati_17, petrescu_17, strinati_17, greschner_17}. 
Flux ladders are successfully realized in various quantum-gas experiments, including real-space~\cite{atala_14, tai_17} and synthetic-dimension implementations~\cite{celi_14, stuhl_15, mancini_15, livi_16, kolkowitz_17, han_19, genkina_19}.
Moreover, they are the most simple models enabling microscopic studies of the Hall response in strongly correlated quantum phases~\cite{prelovsek_99, zotos_00, greschner_19, filippone_19, buser_21}.
%

%
In this paper, we present a snapshot based study of the Hall response in an interacting bosonic flux-ladder model, mimicking actual experiments with quantum-gas microscopes for ultracold atoms in optical lattices~\cite{tai_17}.
The focus is on the measurement of the Hall polarization $P_\mathrm{H}$ and the Hall voltage $V_\mathrm{H}$, which can be probed in the transient dynamics induced by experimentally feasible quench protocols~\cite{buser_21}.
To this end, we draw independent snapshots from matrix-product states according to the perfect sampling scheme outlined by Ferris and Vidal in Ref.~\cite{ferris_12}. 
The employed sampling approach and its peculiarities are discussed in detail. 
For our study it is crucial that the sampling algorithm preserves the $U(1)$ symmetry corresponding to the particle-number conservation of the flux-ladder model.
Moreover, the definitions of the Hall polarization $P_\mathrm{H}$ and the Hall voltage $V_\mathrm{H}$ are based on particle-density distributions as well as on particle currents.
The latter cannot be directly inferred from snapshots taken in the standard Fock measurement basis.
Thus, we employ suitable unitary two-site transformations which enable the sampling of local particle-current distributions.
The current-sampling strategy is exemplified for the case of the vortex-lattice\textsubscript{1/2} phase of interacting bosons on a flux ladder. 
Finally, using a suitable pattern of two-site current transformations, we simultaneously sample the transverse particle-density gradient and the longitudinal particle current required for the estimation of the Hall response in flux ladders from the same snapshot data.
In a broader perspective, our work motivates the feasibility of time-dependent protocols for measurements of the Hall response in optical lattice experiments with quantum-gas microscopes by focusing on actual snapshot data. 
%

%
This paper is organized as follows.
In Sec.~\ref{sec:model}, we introduce the paradigmatic bosonic two-leg flux-ladder model and the ground-state phases which are considered in this paper.
In Sec.~\ref{sec:snapshots}, we present the matrix-product-state based snapshot-sampling approach for generic measurement setups.
Moreover, we discuss how local particle currents can be effectively sampled by means of unitary two-site transformations.
We exemplify the sampling of particle-current statistics in Sec.~\ref{sec:current_snapshots}, focusing on characteristic current patterns in the vortex-lattice\textsubscript{1/2} phase of the flux-ladder model.
In Sec.~\ref{sec:hall_response}, we discuss and exemplify a realistic scheme for the snapshot-based estimation of the Hall response in flux ladders. 
We present and verify snapshot results for the Hall polarization and for the Hall voltage, which are obtained in the framework of quench protocols.
Finally, we summarize our work in Sec.~\ref{sec:summary}.
\section{Bosonic flux-ladder model}
\label{sec:model}
\begin{figure*}
	\centering
	\includegraphics{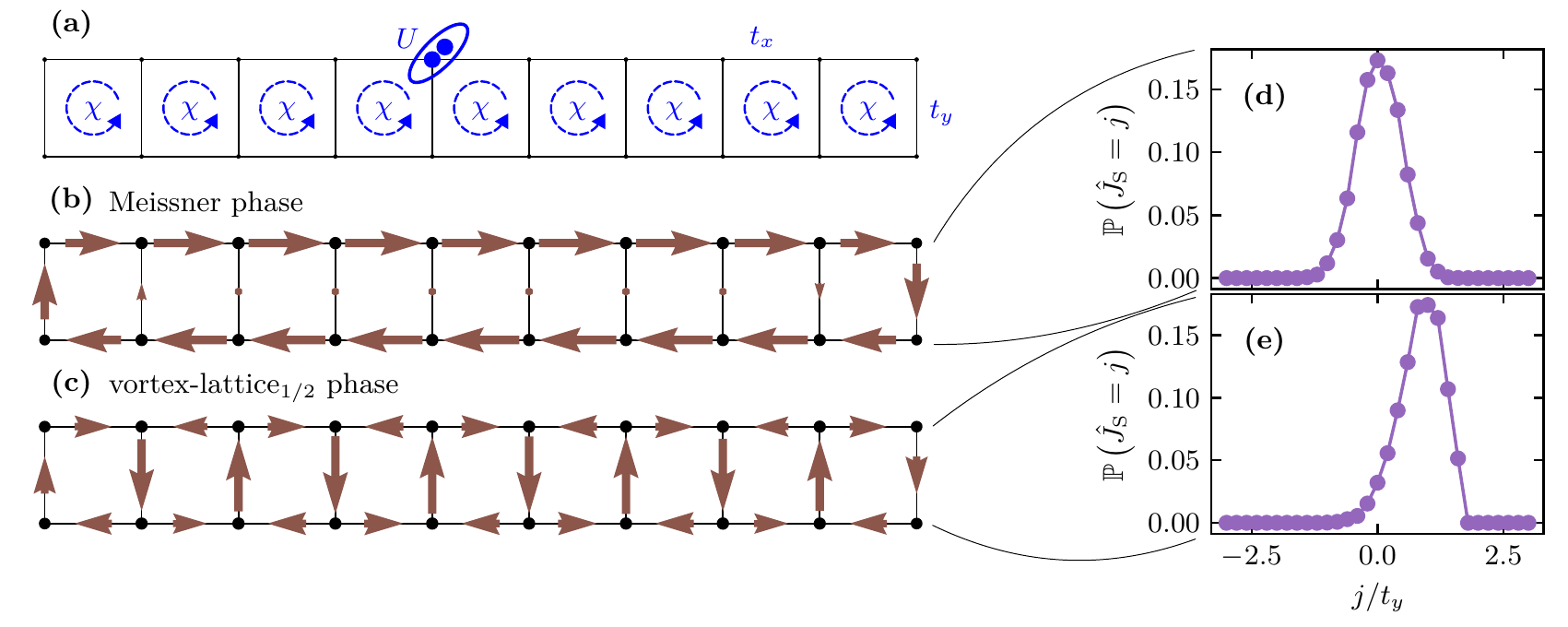}
	\caption{
		(a)~Sketch of the flux-ladder model.
		The Hamiltonian parameters $U$, $\chi$, $t_x$, and $t_y$ are introduced in the context of Eq.~\eqref{eq:fluxladderhamiltonian}.
		The ground-state phases considered in this paper are (b)~the superfluid Meissner phase and (c)~the superfluid vortex-lattice\textsubscript{1/2} phase~\cite{orignac_01}.
		The length of the arrows depicts the strength of the particle currents in a finite-size ladder comprising $L=10$ rungs.
		(d)~Snapshots taken in suitable measurement bases immediately give rise to full counting statistics of nontrivial operators. 
		Here, we show the sampled probability distribution $\mathbb{P}(\hat{J}_\mathrm{S})$ of staggered rung-current operator $\hat{J}$, which is defined in Sec.~\ref{sec:model}, using $N=10^4$ snapshots in the Meissner phase.
		(e)~The probability distribution of the staggered rung-current operator  $\mathbb{P}(\hat{J}_\mathrm{S})$ obtained by means of $N=10^4$ snapshots in the vortex-lattice\textsubscript{1/2} phase.
		In the Meissner phase, the distribution $\mathbb{P}(\hat{J}_\mathrm{S})$ is symmetrically centered around zero, which is in accordance with the expected vanishing of the rung currents, $j^\perp_r=0$.
		In contrast, in the vortex-lattice\textsubscript{1/2} phase, the expectation value of the staggered rung-current operator $\hat{J}_\mathrm{S}$ takes on a finite value.
	} 
	\label{fig:sketch_ladder}
\end{figure*}
In terms of site-local bosonic (annihilation) creation operators $\hat{a}_{r,m}^{(\dagger)}$, the paradigmatic two-leg flux-ladder Hamiltonian reads
\begin{align}
	\hat{H} =
	&- t_{x}\sum_{m=0}^{1}\sum_{r=0}^{L-1} \left(e^{i\left(m-1/2\right)\chi} \hat{a}_{r,m}^\dagger \hat{a}_{r+1,m} + \text{H.c.}\right) \nonumber\\
	&-t_{y}\sum_{r=0}^{L-1} \left(\hat{a}_{r,0}^\dagger \hat{a}_{r,1}+\text{H.c.} \right)+\hat{H}_\mathrm{int}\label{eq:fluxladderhamiltonian}\,,
\end{align}
where $m=0$ and  $m=1$ refer to the lower and upper leg of the ladder, respectively, and $r=0,1,\dots,L-1$ denotes the rung of the ladder.
Particle hopping along the legs and rungs of the ladder is governed by $t_x$ and $t_y$, respectively, and the leg-hopping terms are accompanied by complex phase factors realizing a magnetic flux $\chi$ per plaquette, as shown in Fig.~\ref{fig:sketch_ladder}(a).
The interacting part of the Hamiltonian~\eqref{eq:fluxladderhamiltonian} is explicitly given by $\hat{H}_\mathrm{int} = \frac{U}{2} \sum_{m,r} \hat{n}_{r,m}\left( \hat{n}_{r,m}-1\right)$, with $\hat{n}_{r,m}=\hat{a}^\dagger_{r,m}\hat{a}_{r,m}$, and the interparticle interaction strength is parametrized by $U$.
%
%
The flux-ladder Hamiltonian~\eqref{eq:fluxladderhamiltonian} commutes with $\hat{n}_\mathrm{tot}= \sum_m\sum_r  \hat{n}_{r,m} $ and, thus, the total number of particles is conserved.
Throughout this paper, we refer to the particle filling, meaning the total number of particles divided by the total number of lattice sites, with $\nu= \left\langle \hat{n}_\mathrm{tot} \right\rangle / (2L)$.
Note that here and in the following, angled brackets denote expectation values.
%

%
Because of the interplay between effective magnetic fields and interactions among quantum particles, flux ladders exhibit various ground-state phases~\cite{orignac_01, greschner_16}.
The list of accessible ground states includes biased-ladder states~\cite{wei_14}, charge-density waves~\cite{greschner_17}, as well as precursors of fractional quantum Hall states~\cite{grusdt_14, petrescu_15, cornfeld_15, petrescu_17, strinati_17, strinati_19}.
In the following, we consider model parameters corresponding to the superfluid Meissner phase~\cite{orignac_01, piraud_15} and to the superfluid vortex-lattice\textsubscript{1/2} phase~\cite{dhar_12, dhar_13, greschner_15}, noting that Mott-insulating variants of both ground-state phases can be stabilized for commensurable particle fillings.
Ground states in the Meissner phase exhibit homogeneous particle-density profiles and uniform particle currents running along the legs of the ladder in opposite directions, while rung currents are vanishing in the center of the system.
They adiabatically extend to the noninteracting regime, corresponding to $U=0$~\cite{huegel_14}. 
Figure~\ref{fig:sketch_ladder}(b) shows a ground state in the Meissner phase, which is realized for $U/t_x = 2$, $t_y/t_x = 1.6$, $\chi/\pi = 0.2$, and $\nu=0.8$.
The key feature of vortex-lattice phases are localized current vortices. 
In contrast to the Meissner phase, vortex-lattice phases require weak but finite interparticle interactions.
Figure~\ref{fig:sketch_ladder}(c) shows the vortex-lattice\textsubscript{1/2} phase, which will be discussed later on, with an alternating pattern of rung currents and a unit cell comprising two plaquettes of the ladder.
The model parameters considered Fig.~\ref{fig:sketch_ladder}(c) are the same as in Fig.~\ref{fig:sketch_ladder}(b), except for $\chi/\pi = 0.98$.
%

%
In addition to particle-density profiles $\left\langle \hat{n}_{r,m} \right\rangle$, local particle-currents are key for the characterization of the various ground-state phases of the flux-ladder model.
From the continuity equation for the occupation of individual lattice sites, local particle currents along the rungs and legs of the ladder are found to be given by $j^\perp_{r} = -i t_{y} \langle \hat{a}_{r,0}^\dagger \hat{a}_{r,1} \rangle +\text{H.c.} $ and ${j^\parallel_{r,m} = - i t_x e^{i\left(m-1/2\right)\chi} \langle \hat{a}_{r,m}^\dagger \hat{a}_{r+1,m} \rangle + \text{H.c.}}$, respectively.
Moreover, the staggered rung-current operator is defined as $\hat{J}_\mathrm{S} = -i t_{y} \sum_r (-1)^r \hat{a}_{r,0}^\dagger \hat{a}_{r,1}/L + \text{H.c.}$.
%

%
The probability distribution of the staggered rung-current operator $\mathbb{P}(\hat{J}_\mathrm{S})$ is shown in Fig.~\ref{fig:sketch_ladder}(d) and in Fig.~\ref{fig:sketch_ladder}(e) for the Meissner phase and for the vortex-lattice\textsubscript{1/2} phase, respectively.
To calculate it, we used snapshots of the particle currents, which we will discuss in great detail in this paper.
The distribution $\mathbb{P}(\hat{J}_\mathrm{S})$ shown in Fig.~\ref{fig:sketch_ladder}(d) is symmetrically centered around zero, which is in accordance with the vanishing rung currents in the Meissner phase.
In the vortex-lattice\textsubscript{1/2} phase, the pattern of alternating rung currents gives rise to a finite expectation value of the staggered rung-current operator, which can be inferred from the distribution $\mathbb{P}(\hat{J}_\mathrm{S})$ shown in Fig.~\ref{fig:sketch_ladder}(e).
\section{Drawing snapshots from matrix-product states}
\label{sec:snapshots}
In the following, we first account for a generic quantum measurement setup in Sec.~\ref{sec:snapshots:measurementsetup}.
Second, we discuss how snapshots can be sampled from matrix-product states in Sec.~\ref{sec:snapshots:mpsapproach}.
In Sec.~\ref{sec:snapshots:currents}, we focus on the sampling of particle currents in flux-lattice models using suitable unitary two-site transformations.
\subsection{Quantum measurement setup}
\label{sec:snapshots:measurementsetup}
Here, we consider generic quantum lattice models with a total number of $L$ sites, labeled by $i=1,2,\dots,L$. 
For simplicity, we assume all sites to be of the same kind, with a site-local Hilbert space spanned by $d$ basis states ${\ket{e_i}=\ket{1}, \ket{2},\dots,\ket{d}}$. 
A generic quantum state of interest is denoted by $\ket{\psi}$ and its matrix-product-state representation takes the form
\begin{align}
	\ket{\psi} = \sum\limits_{e_1 = 1}^{d}\sum\limits_{e_2 = 1}^{d}\dots\sum\limits_{e_L = 1}^{d}
	&M_{1}^{e_1}M_{2}^{e_2}\dots M_{L}^{e_L}\nonumber\\ 
	&\times \ket{e_1,e_2,\dots,e_L} \label{eq:mps_psi}\,,
\end{align}
with matrices $M_{i}^{e_i}$.
%

%
Snapshots are obtained from a simultaneous measurement of site-($i$)-local observables, given by ${\hat{O}_i = \sum_{k_i} o_{k_i} \hat{P}_{k_i}}$.
Here, the $o_{k_i}$ represent site-local measurement outcomes corresponding to site-local projectors $\hat{P}_{k_i}$.
For different lattice sites, the site-local observables do not necessarily need to coincide.
However, the site-local projectors need to add up to the site-local identity operator, $\sum_{k_i} \hat{P}_{k_i}=\mathbb{I}_i$.
While taking a snapshot, all site-local measurements are performed at the same time.
Thus, the global snapshot observable takes the form
\begin{align}
	\hat{O} = \sum\limits_{k_1}\sum\limits_{k_2}\dots\sum\limits_{k_L}
	& \left(o_{k_1}, o_{k_2}, \dots, o_{k_L}\right)\nonumber\\ 
	&\times \hat{P}_{k_1} \hat{P}_{k_2}\dots \hat{P}_{k_L}\,,
\end{align}
where the tuples $\left(o_{k_1}, o_{k_1}, \dots, o_{k_L}\right)$ represent the global measurement outcomes, that is, the snapshots of interest.
Most often, the Fock basis consisting of site-local occupation numbers $n_i$ is used, but the formalism is more general.
In this article, we will also consider particle-current measurement bases.
\subsection{Matrix-product-state approach}
\label{sec:snapshots:mpsapproach}
\begin{figure}
	\centering
	\includegraphics{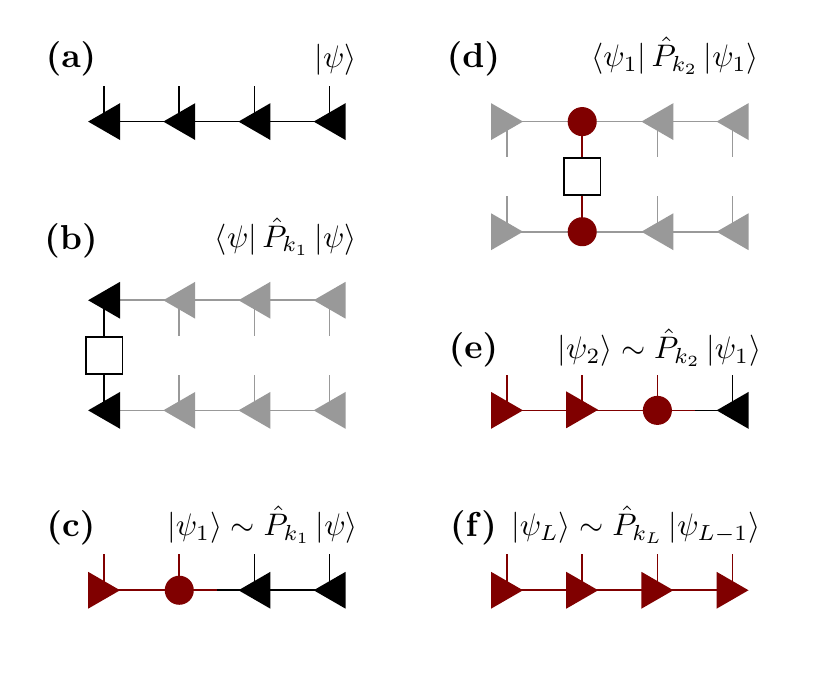}
	\caption{
		Sketch of the snapshot sampling routine~\cite{ferris_12}. 
		In order to draw independent snapshots, we proceed as follows.
		(a)~Starting off with a copy of the matrix-product-state representation of the underlying quantum state of interest $\ket{\psi}$, which is comprised of right-canonical tensors (left triangles)~\cite{schollwoeck_11}, the probability of finding a certain measurement outcome $o_{k_1}$ on the first site, $\mathbb{P}(o_{k_1})$, can be directly evaluated from the first (leftmost) tensor, as shown in~(b).
		In accordance with the probability distribution $\mathbb{P}\left(o_{k_1}\right)$, a local projector $P_{k_1}$ is picked at random, determining the first digit of the snapshot $o_{k_1}$.
		(c)~A transformed matrix-product state $\ket{\psi_1} = P_{k_1}\ket{\psi}/\sqrt{\bra{\psi}P_{k_1}\ket{\psi}}$ is obtained by locally applying $P_{k_1}$ and renormalizing the projected state while shifting the center of orthogonality to the second site (red circle) by means of a singular value decomposition~\cite{schollwoeck_11}.
		As compared to the original state $\ket{\psi}$, the first and the second tensor of $\ket{\psi_1}$ are transformed, which is indicated by the different colors.
		The conditional probabilities $\mathbb{P}\left(o_{k_2} | o_{k_1} \right)$ are determined efficiently from $\ket{\psi_1}$, as shown in (d).
		(e)~After drawing the second digit of the snapshot, $o_{k_2}$, at random and in accordance with $\mathbb{P}\left(o_{k_2} | o_{k_1} \right)$, the corresponding projector $P_{k_2}$ is applied to $\ket{\psi_1}$.
		The projected state is subsequently normalized while shifting the center of orthogonality to the right, giving rise to $\ket{\psi_2}$. 
		(f)~Repeating these steps in a complete left-to-right sweep eventually gives rise to a left-canonical matrix-product state $\ket{\psi_L}$, indicated by the right-triangular shaped tensors, and the desired snapshot $\left(o_{k_1}, o_{k_1}, \dots, o_{k_L}\right)$ sampled from $\ket{\psi}$.
	} 
	\label{fig:mps_sampling}
\end{figure}

%
%
%
%
%
%
%
%
%
%
In order to draw independent snapshots from matrix-product states, we essentially follow the perfect sampling approach outlined by Ferris and Vidal in Ref.~\cite{ferris_12}. 
Here, perfect sampling refers to the fact that this scheme generates perfectly uncorrelated snapshots, without the need to account for additional equilibration or autocorrelation times, as in Markov chain Monte Carlo approaches~\cite{sandvik_07, schuch_08, grusdt_19}.
For our results concerning the bosonic flux-ladder model, that will be presented later on, it is crucial to additionally account for the $U(1)$ symmetry corresponding to the conservation of the particle number while generating snapshots from matrix-product states.
Matrix-product states as well as their canonical forms and elementary operations, which are taken for granted in the following, are discussed in detail in Ref.~\cite{schollwoeck_11}.
We emphasize that for the snapshot sampling of matrix-product states, the evaluation of conditional probabilities $\mathbb{P}\left(o_{k_i}| o_{k_{i-1}}, o_{k_{i-2}}, \dots, o_{k_1}\right)$ and the application of site-local projectors $P_{k_i}$ play a central role.
A sketch illustrating the sampling approach employed in this work is shown in Fig.~\ref{fig:mps_sampling}.
%

%
Given a right-canonical matrix-product-state representation of the underlying wave function of interest $\ket{\psi}$, snapshots are computed by sweeping through the tensors from left to right, that is, from $i=1$ to $i=L$, as shown in Fig.~\ref{fig:mps_sampling}.
First, the probabilities of the possible measurement outcomes $\mathbb{P}\left(o_{k_1}\right)$ on the first ($i=1$) site can be directly evaluated because of the right-canonical form of the matrix-product state.
In accordance with the probability distribution $\mathbb{P}\left(o_{k_1}\right)$, a local projector $\hat{P}_{k_1}$ is chosen at random, determining the first digit of the snapshot, that is, $o_{k_1}$.
Next, the local projector $\hat{P}_{k_1}$ is applied to $\ket{\psi}$. 
On the matrix-product-state level this is a local operation.
The projected matrix-product state is subsequently normalized by means of a singular-value decomposition, shifting the center of orthogonality to the second ($i=2$) site.
The so transformed state is denoted as $\ket{\psi_1}$,
\begin{align}
	\ket{\psi_1} = \hat{P}_{k_1} \ket{\psi}/\sqrt{\bra{\psi}\hat{P}_{k_1}\ket{\psi}}\,.
\end{align}
Note that the steps so far only involved transformations of the first and the second of the matrix-product-state tensors, that is, using the notation from Eq.~\eqref{eq:mps_psi},  $M_1$ and $M_2$.
%

%
From $\ket{\psi_1}$, the conditional probabilities of the possible measurement outcomes on the second site $\mathbb{P}\left(o_{k_2}| o_{k_{1}}\right)$ can be directly evaluated from local contractions because of the mixed-canonical form. 
Hence, the second digit of the snapshot $o_{k_2}$ is drawn at random, in accordance with $\mathbb{P}\left(o_{k_2}| o_{k_{1}}\right)$.
The matrix-product state is then projected by applying $\hat{P}_{k_2}$ locally and normalized while shifting the center of orthogonality further to the right.
Repeating these steps in a complete left-to-right sweep, according to $\ket{\psi_n} = \hat{P}_{k_n} \ket{\psi_{n-1}}/\sqrt{\bra{\psi}\hat{P}_{k_n}\ket{\psi_{n-1}}}$, while maintaining a canonical form of the transformed state eventually yields the snapshot $\left(o_{k_1}, o_{k_1}, \dots, o_{k_L}\right)$.
Note that the protocol discussed here can be understood as an efficient evaluation of the conditional probabilities on right-hand side of the following identity:
\begin{align}
	\mathbb{P}\left(o_{k_1} o_{k_{2}}, \dots, o_{k_L}\right)=
	&\mathbb{P}\left(o_{k_1}\right)
	\mathbb{P}\left(o_{k_2} | o_{k_1}\right)\times\dots\\
	&\dots \mathbb{P}\left(o_{k_L} | o_{k_{L-1}}, o_{k_{L-2}},\dots,o_{k_{1}}\right)\nonumber\,.
\end{align}
%

%
Moreover, the sampling protocol discussed above is applicable to matrix-product states with protected quantum symmetries, such as the $U(1)$ symmetry corresponding to the particle number conservation of the flux-ladder Hamiltonian~\eqref{eq:fluxladderhamiltonian}.
For this, it needs to be ensured that the successive projections, which are applied during the left-to-right sweep, preserve the symmetry-sector information on the level of the matrix-product state~\cite{hubig_17}.   
\subsection{Particle currents}
\label{sec:snapshots:currents}
Next, we demonstrate that we can take snapshots in a measurement basis different from the standard Fock basis.
Specifically, we show how snapshots of particle currents can be obtained.
Later on, this will become important for the estimation of the Hall response.
There, one requires information about particle densities as well as about particle currents.
Thus, we introduce suitable unitary two-site transformations, which allow to simultaneously determine both, the required particle-density profiles and particle-current profiles, from the same snapshot data.
%

%
In order to introduce the unitary two-site current transformation, we consider the generic case of two neighboring bosonic lattice sites, labeled by indices $k$ and $l$.
Particle exchange between these lattice sites is assumed to be governed by a generic complex hopping term $\hat{T}_{k,l}$,
\begin{align}
	\hat{T}_{k,l} = t_{k,l}\exp(i\phi_{k,l})\hat{a}_{k}^\dagger \hat{a}_l + \text{h.c.}\,,	
\end{align}
with bosonic annihilation (creation) operators $\hat{a}_{\alpha}^{(\dagger)}$ (for $\alpha=k,l$) and real-valued $t_{k,l}$ and $\phi_{k,l}$.
The corresponding particle current operator, which is derived from the continuity equation for the occupation of individual lattice sites, reads
\begin{align}
	\hat{J}_{k,l} = -i t_{k,l}\exp(i\phi_{k,l})\hat{a}_{k}^\dagger \hat{a}_l + \text{h.c.}\,.
\end{align}
%

%
For the occupation number operators $\hat{n}_{\alpha} = \hat{a}_{\alpha}^\dagger \hat{a}_\alpha$ of nearest-neighbor lattice sites $(\alpha=k,l)$, the unitary two-site current transformation is given by $\tilde{n}_{\alpha} = \hat{U}_{k,l} \hat{n}_{\alpha} \hat{U}_{k,l}^\dagger$, with
\begin{align}
	\hat{U}_{k,l} = \exp\left(i\frac{\pi}{4} \left( e^{i \phi_{k,l}} \hat{a}_{k}^\dagger \hat{a}_l + \mathrm{H.c.}\right) \right)\label{eq:current_transform}\,.
\end{align}
Importantly, the so transformed operators $\tilde{n}_{\alpha}$ satisfy
\begin{align}
	\tilde{n}_k + \tilde{n}_l & = \hat{n}_k + \hat{n}_l\,,
	&\tilde{n}_k - \tilde{n}_l & = \hat{J}_{k,l}/t_{k,l}\label{eq:ntilde_props}\,.
\end{align}
Hence, sampling the particle density profiles of two current-transformed lattice sites, $\tilde{n}_k$ and $\tilde{n}_l$, directly gives rise to the the statistics of their joint occupation number $\hat{n}_k + \hat{n}_l$ as well as the to the statistics of the local particle current $\hat{J}_{k,l}$ between them.
\section{Sampling particle currents in bosonic flux ladders}
\label{sec:current_snapshots}
In this section, we exemplify the snapshot sampling of local particle currents using matrix-product states.
To this end, we consider the superfluid vortex-lattice\textsubscript{1/2} phase of the paradigmatic bosonic flux-ladder model~\cite{orignac_01}. 
Here, a vortex-lattice\textsubscript{1/2} matrix-product state is obtained by means of a density-matrix renormalization-group simulation, considering an interparticle interaction strength $U/t_x=2$, a rung-hopping strength $t_y/t_x=1.6$, a magnetic flux $\chi/\pi=0.98$, and a particle filling of $\nu=0.8$ bosons per lattice site~\cite{greschner_16}.
Localized current vortices on every second plaquette of the ladder, which give rise to a regular and alternating pattern of rung currents, are a key feature of the vortex-lattice\textsubscript{1/2} phase; see Fig.~\ref{fig:sketch_ladder}(c).
%

%
\begin{figure*}
	\centering
	\includegraphics{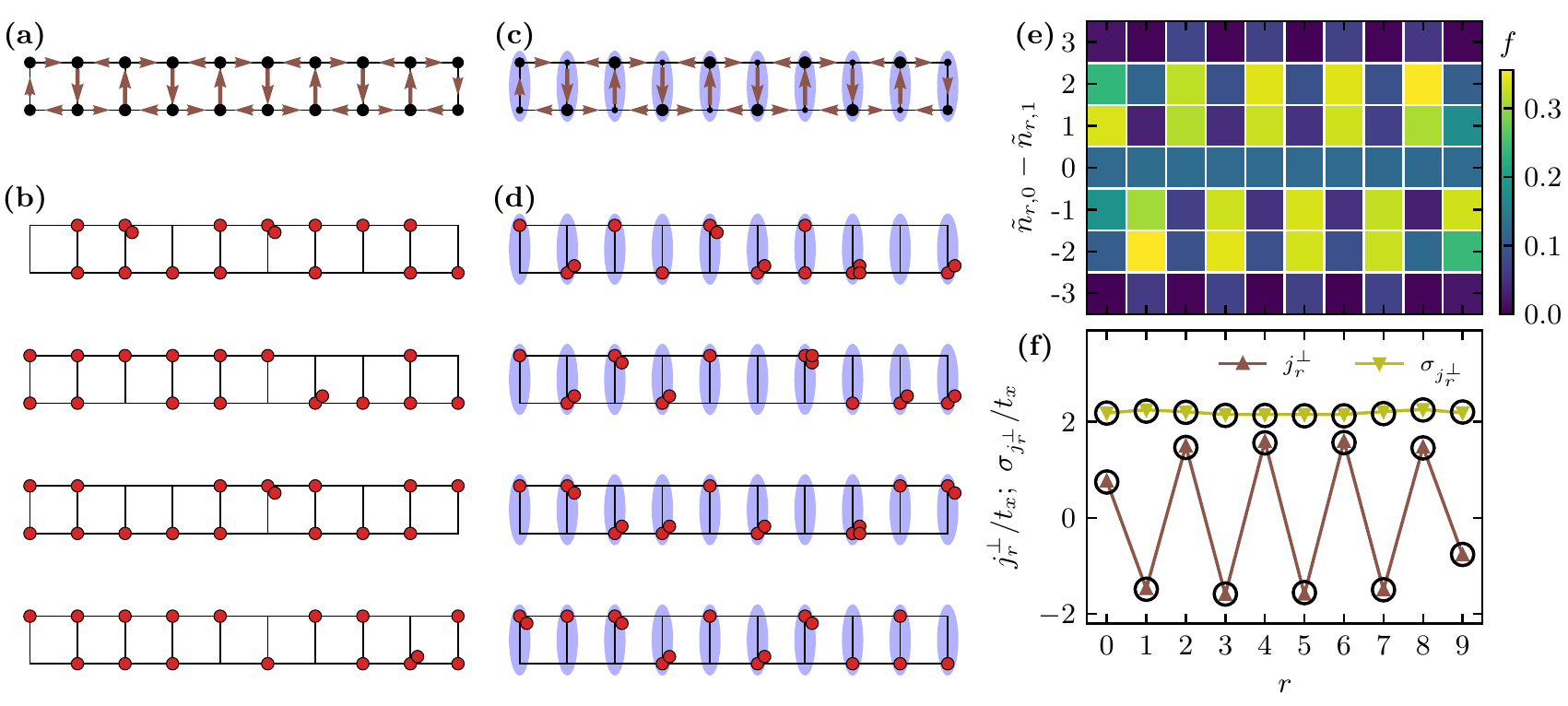}
	\caption{
		Particle-current sampling from vortex-lattice\textsubscript{1/2} snapshots of a bosonic flux ladder comprising ten rungs.
		Considering $U/t_x = 2$, $t_y/t_x = 1.6$, $\chi/\pi=0.98$, $\nu=0.8$, and a local bosonic cutoff at at most three particles per lattice site. 
		(a)~Vortex-lattice\textsubscript{1/2} phase with local particle currents and local particle densities, which are directly computed from the matrix-product ground-state representation.
		Red arrows indicate the direction and, by their length, the strength of local particle currents. 
		The size of the black circles indicates the local particle density, which is homogeneous in the vortex-lattice\textsubscript{1/2} phase. 
		(b)~Four independent ground-state snapshots corresponding to the standard site-local Fock-measurement basis.
		From these Fock-basis snapshots, the site-local particle density can be reconstructed. 
		(c)~Two-site unitary transformations, indicated by the blue ellipses and introduced in the context of Eq.~\eqref{eq:current_transform}, are applied to neighboring sites sharing a rung of the ladder.
		Hence, these snapshots taken in the $\tilde{n}_{r,m}$ basis directly reveal the local rung currents.
		(d)~Four independent ground-state snapshots corresponding to the $\tilde{n}_{r,m}$-measurement basis.
		(e)~Probability distribution $f$ of $(\tilde{n}_{r,0}-\tilde{n}_{r,1})$ obtained from $N=10^4$ snapshots.
		Note the central symmetry between the rungs $r=4$ and $r=5$ of the ladder.
		(f)~Average local rung currents $j^\perp_r$ obtained from the sampled $(\tilde{n}_{r,0}-\tilde{n}_{r,1})$ distribution (brown up-triangles) and the corresponding standard deviations $\sigma_{j^\perp_r}$ (yellow down-triangles).
		Lines are a guide to the eye.
		Open circles show results that are directly computed from the underlying matrix-product state.
	} 
	\label{fig:current_statistics}
\end{figure*}
Figure~\ref{fig:current_statistics}(a) shows the particle-current profile and the homogeneous particle-density profile in the superfluid vortex-lattice\textsubscript{1/2} phase of a ladder comprising ten rungs.
Sampling snapshots in the standard site-local Fock-measurement basis immediately give rise to the statistics of $\hat{n}_{r,m}$ and, thus, to the average particle-density profile $\left\langle \hat{n}_{r,m} \right\rangle$.
Four independent snapshots are shown in Fig.~\ref{fig:current_statistics}(b).
They are drawn from the vortex-lattice\textsubscript{1/2} matrix-product state considering the standard Fock basis.
%

%
In order to resolve characteristic particle-current patterns of the vortex-lattice\textsubscript{1/2} phase, we apply the unitary two-site current transformation, introduced in the context of Eq.~\eqref{eq:current_transform}, to neighboring lattice sites on each rung of the ladder, as indicated by the blue ellipses in Fig.~\ref{fig:current_statistics}(c).
Sampling the site-local occupations of the so transformed state directly give rise to the distributions of $\tilde{n}_{r,m}$. 
Moreover, according to Eq.~\eqref{eq:ntilde_props}, the rungwise differences in the occupation of the transformed lattice sites $(\tilde{n}_{r,0}-\tilde{n}_{r,1})$ yield the statistics of the local rung currents $j_r^\perp$.  
As an example, Fig.~\ref{fig:current_statistics}(d) shows four independent snapshots drawn from the underlying vortex-lattice\textsubscript{1/2} matrix-product state considering the $\tilde{n}_{r,m}$ basis.
%

%
For a local bosonic cutoff at at most three particles per lattice site, the distribution of $(\tilde{n}_{r,0}-\tilde{n}_{r,1})$ obtained from $N=10^4$ snaphsots is shown separately for each rung $r$  in Fig.~\ref{fig:current_statistics}(e).
It possesses a point symmetry between the central rungs $r=4$ and $r=5$ of the ladder. 
Most importantly, the maximum points of $(\tilde{n}_{r,0}-\tilde{n}_{r,1})$ alternate in sign between neighboring rungs $r$, which is a clear fingerprint of the vortex-lattice\textsubscript{1/2} phase.
%

%
In Fig.~\ref{fig:current_statistics}(f), brown up-triangles show the average local rung currents $j_r^\perp$ calculated from the snapshot-sampled ${(\tilde{n}_{r,0}-\tilde{n}_{r,1})}$ distribution.
Note that the lines are a guide to the eye and that alternating rung currents are expected for the vortex-lattice\textsubscript{1/2} phase.
The snapshot-sampled average currents $j_r^\perp$ perfectly coincide with the values that are directly calculated from the underlying vortex-lattice\textsubscript{1/2} state and that are indicated by the black open circles.
Additionally, yellow down-triangles show standard deviations of the rung currents $\sigma_{j_r^\perp}$, as obtained from the snapshot-sampled ${(\tilde{n}_{r,0}-\tilde{n}_{r,1})}$ distribution shown in Fig.~\ref{fig:current_statistics}(e).
The sampled results for $\sigma_{j_r^\perp}$ also perfectly coincide with the standard deviations that are directly computed from the underlying matrix-product state, indicated by the black open circles.
%

%
Here, it is worth noting that snapshots taken in the $\tilde{n}_{r,m}$ basis, as described above and shown in Fig.~\ref{fig:current_statistics}(d), also give rise to the full-counting statistics of non-trivial observables, such as the staggered rung-current operator $\hat{J}_\mathrm{S}$, introduced in Sec.~\ref{sec:model}. 
Indeed, the probability distributions $\mathbb{P}(\hat{J}_\mathrm{S})$ corresponding to the Meissner phase and to the vortex-lattice\textsubscript{1/2} phase that are shown in Fig.~\ref{fig:sketch_ladder}(d) and in Fig.~\ref{fig:sketch_ladder}(e), respectively, are obtained from $N=10^4$ snapshots.
%

%
To sum up, Fig.~\ref{fig:current_statistics} shows that the statistics of local particle currents in the vortex-lattice\textsubscript{1/2} phase can be effectively sampled using unitary two-site transformations on the matrix-product-state level.
While standard snapshots, corresponding to Fock-basis measurements, show a homogeneous particle-density profile, current snapshots, which are taken after the rungwise unitary transformation of the underlying ground-state, clearly reveal the structure of the underlying vortex-lattice\textsubscript{1/2} phase.
\section{Sampling the Hall response of flux ladders}
\label{sec:hall_response}
In this section, we present a snapshot based analysis of the Hall response in the Meissner phase, which is the most prominent ground-state phase of the flux-ladder model.
Concretely, we focus on the Hall polarization $P_\mathrm{H}$ and on the Hall voltage $V_\mathrm{H}$~\cite{buser_21}.
We first recap the definition of the Hall polarization $P_\mathrm{H}$ and the Hall Voltage $V_\mathrm{H}$ in the paradigmatic two-leg flux-ladder model~\eqref{eq:fluxladderhamiltonian} and how both quantities can be probed by means of realistic quench protocols in Sec.~\ref{sec:hall_response:definitions}.
In Sec.~\ref{sec:hall_response:snapshot_approach}, we put forward a suitable pattern of unitary two-site current transformations, enabling the direct estimation of the Hall polarization $P_\mathrm{H}$ from snapshot data.
In Sec.~\ref{sec:hall_response:static_tilt}, we present snapshot results from time-dependent quench simulations.
Results concerning the Hall voltage $V_\mathrm{H}$ are discussed in Sec.~\ref{sec:hall_response:hall_voltage}.
%
\subsection{Definition of the Hall response}
\label{sec:hall_response:definitions}
At the core of the characterization of the ground-state Hall response, meaning the Hall polarization $P_\mathrm{H}$ and the Hall voltage $V_\mathrm{H}$, are the transverse polarization $p_y$ and the longitudinal current $j_x$.
For the two-leg ladder Hamiltonian~\eqref{eq:fluxladderhamiltonian}, both quantities are explicitly given by
\begin{align}
	p_y &=\frac{1}{2L} \left\langle \hat{P}_y \right\rangle\,,&
	j_x &= \frac{1}{2L} \sum_{m=0}^{1} \sum_{r=0}^{L-1} j_{r,m}^{\parallel}\,,
\end{align}
with $\hat{P}_y = \sum_m \sum_r \left( m -1/2 \right) \hat{n}_{r,m} $ and local leg currents $j_{r,m}^{\parallel}$ as defined in Sec.~\ref{sec:model}.
%

%
The Hall polarization is defined as the ratio between the transverse polarization $p_y$ and the longitudinal current $j_x$,
\begin{align}
	P_\mathrm{H} = p_y / j_x\,.
\end{align}
In a ring-shaped ladder with periodic boundary conditions, a finite longitudinal current $j_x$ is typically induced by means of an additional Aharonov-Bohm flux piercing the ring~\cite{prelovsek_99, zotos_00, greschner_19, filippone_19}. 
In systems with open boundaries, transient longitudinal currents can be realized by means of time-dependent ramps or tilts.
Note that the consistent calculation of the Hall polarization $P_\mathrm{H}$ and the Hall voltage $V_\mathrm{H}$ in time-dependent protocols for systems with open boundaries and in the ground states of ring-shaped ladders is discussed in detail in Ref.~\cite{buser_21}.
%

%
The formal definition of the Hall voltage $V_\mathrm{H}$ requires the extension of the flux-ladder Hamiltonian by an additional transverse potential, $\hat{H} + \mu_y \hat{P}_y$.
Concretely, in a ring-shaped ladder with periodic boundary conditions, $\mu_y$ needs to be adjusted in such a way that the transverse polarization $p_y$ vanishes in the  ground state.
For this suitably chosen value of $\mu_y$, the Hall voltage is then defined as
\begin{align}
	V_\mathrm{H} = \mu_y / j_x\,.
\end{align}
However, in Ref.~\cite{buser_21} it is also shown that in Meissner and vortex-lattice phases, the Hall voltage $V_\mathrm{H}$ can be faithfully approximated by means of $V_\mathrm{H} = P_\mathrm{H}(\mu_y/p_y)$, with $(\mu_y/p_y)$ as obtained from an independent simulation with a small but finite bias, $\mu_y\to 0$.
We recap that the Hall voltage $V_\mathrm{H}$ is a key quantity:
As compared to the Hall polarization $P_\mathrm{H}$, it exhibits a remarkable robustness with respect to the particle filling $\nu$, the interparticle interaction strength $U$, and also for multileg ladders~\cite{buser_21}.
\subsection{Snapshot approach}
\label{sec:hall_response:snapshot_approach}
\begin{figure}
	\centering
	\includegraphics{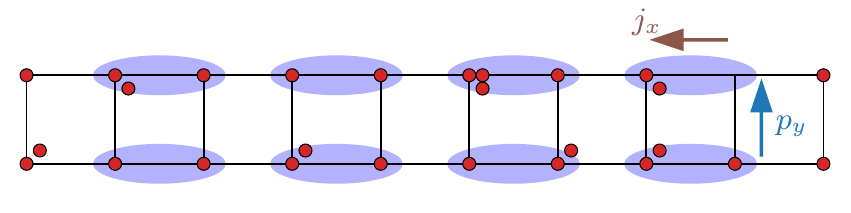}
	\caption{
		Sketch of the current-transformation scheme employed for the estimation of the Hall response.
		Two-site unitary transformations, indicated by the blue ellipses and introduced in the context of Eq.~\eqref{eq:current_transform}, are applied to neighboring lattice sites along the longitudinal direction, as shown above.
		The scheme enables the simultaneous estimation of the longitudinal current $j_x$ and the transverse polarization $p_y$ from the same snapshot data.
		It is discussed in more detail in the main text.
	}
	\label{fig:sketch_current_transformation_hall_response}
\end{figure}
Probing the Hall response in a flux ladder requires measurements of the transverse particle-density gradient and the longitudinal particle current.
Figure~\ref{fig:sketch_current_transformation_hall_response} shows a pattern of unitary two-site current transformations that allow to determine both quantities simultaneously from the same snapshot data.
Concretely, the two-site transformations, which are introduced in the context of Eq.~\eqref{eq:current_transform}, are applied to neighboring lattice sites of the ladder in the longitudinal direction, as indicated by the blue ellipses.
The transverse polarization and the longitudinal current of a (potentially mixed) state $\rho$ of the ladder, that is transformed as $\bar{U}^\dagger_{k,l} \rho \bar{U}_{k,l}$ with $\bar{U} = \prod_{m = 0}^{1}\prod_{r\prime = 0}^{L/2}\hat{U}_{(2r\prime, m), (2r\prime+1, m)}$, can be directly snapshot-sampled via
\begin{align}
	p_y &=  \sum_{m=0}^1 \sum_{r=0}^{L-1} \frac{2m - 1}{4L} \left\langle \tilde{n}_{r,m} \right\rangle \,,\\
	j_x &= \sum_{m=0}^1 \sum_{r\prime=0}^{L/2} \frac{t_x}{2} \left\langle \tilde{n}_{2r\prime,m} - \tilde{n}_{2r\prime+1,m}\right\rangle\,.
\end{align}
%

%
Note that it is sufficient to restrict the unitary current transformations to a central region of the ladder, in which the quantities of interest can be faithfully determined.
For instance, for the results shown in Sec.~\ref{sec:hall_response:static_tilt} only the eight most central rungs of a ladder comprising a total number of $L=40$ rungs are transformed in order to probe the Hall polarization in a suitable quantum quench. 
%

%
In quantum-gas experiments realizing flux ladders with optical lattices~\cite{atala_14} and with a quantum gas microscope~\cite{tai_17}, the required snapshots for measurements of particle currents can be obtained as follows.
First, all sites are decoupled and the particle-hopping dynamics are frozen by deepening the lattice potentials.
Second, an optical super-lattice is used to switch back on tunneling between neighboring lattice sites $k,l$ as shown in Fig.~\ref{fig:sketch_current_transformation_hall_response}.
During this step, interactions between bosons on the same site must be turned off. 
If the latter are realized through a magnetic Feshbach resonance, this can be achieved by tuning the magnetic field; if interactions cannot be tuned efficiently by an external field, the atoms in the individual lattice sites can first be loaded into elongated one-dimensional tubes along the third direction, lowering their density and effectively removing the interactions.
The extra tunneling is switched on for a period of time corresponding to a $\pi/2$-pulse, effectively rotating the bosonic operators to a basis $\hat{a}_{\pm, k,l} = (\hat{a}_k \pm \hat{a}_{l}) / \sqrt{2}$. 
Using the super-lattice again to switch on a staggered potential $+\Delta$ ($-\Delta$) on sites $k$ ($l$) for a controlled period of time $\tau$, allows to rotate the measurement basis to $\tilde{a}_{\pm, k,l} = (\tilde{a}_k \pm i e^{-i \phi_{k,l}} \tilde{a}_{l}) / \sqrt{2}$, where the relative phase $\phi_{k,l} - \pi/2 \propto \Delta \tau$ is controlled by the offset $\Delta$ and the duration $\tau$.
%

%
After applying the above sequence, a measurement of the local occupation numbers on the physical lattice sites $k$ and $l$ yields the densities in the transformed basis, $\tilde{n}_{k,l}$. 
Expressed in the original basis, where the flux-ladder was defined, this reads
\begin{equation}
    \tilde{n}_k = \frac{1}{2} \left( \hat{n}_k + \hat{n}_l -\hat{J}_{k,l} \right),
\end{equation}
and similar but with $-\hat{J}_{k,l}$ for $\tilde{n}_l$. 
Hence, combining these expressions yields the total particle number and the current as described in Eq.~\eqref{eq:ntilde_props} above. 
This demonstrates that the total density and the required currents can efficiently be read of from individual snapshots in a realistic experimental setting. 
\subsection{Snapshot analysis of a static tilt}
\label{sec:hall_response:static_tilt}
\begin{figure}
	\centering
	\includegraphics{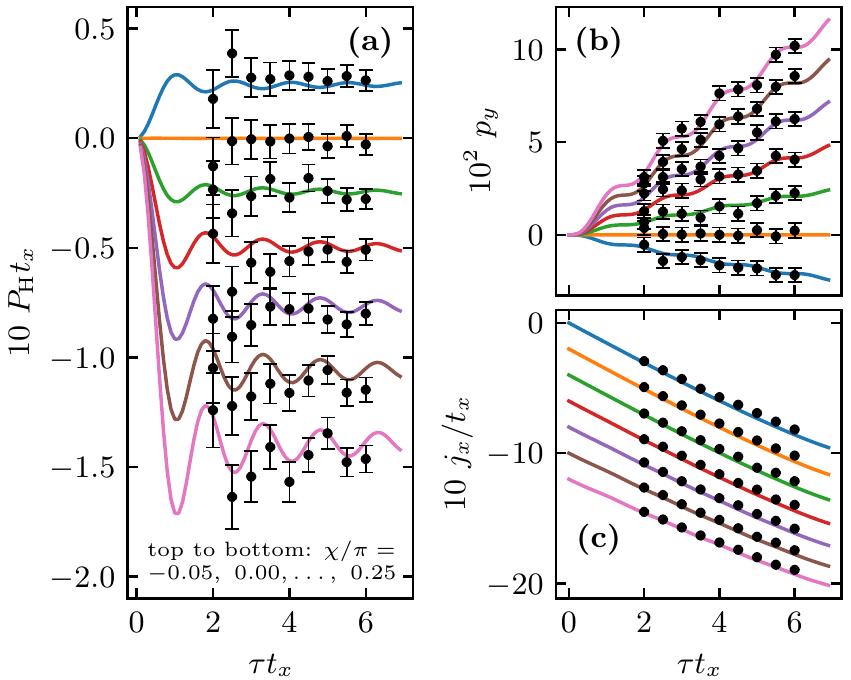}
	\caption{
		Snapshot based estimation of the Hall polarization in a statically tilted two-leg ladder. 
		Considering $U/t_x = 2 $, $t_y/t_x = 1.6$,  $\mu_x/t_x = 0.1$, $\nu=0.8$, and a local bosonic cutoff at at most three particles per lattice site.
		Unitary two-site current transformations, enabling the sampling of the Hall response, are applied to neighboring lattice sites between the eight most central rungs of each leg of the ladder, as described in the context of Fig.~\ref{fig:sketch_current_transformation_hall_response}.
		(a) Transient dynamics of the Hall polarization $P_\mathrm{H} = p_y/j_x$ after statically tilting the ladder in the longitudinal direction.
		The solid lines show exact results that are directly computed from time-evolved matrix-product states, corresponding to different values of the magnetic flux per plaquette $\chi$. 
		From top to bottom, the different lines are for $\chi/\pi = -0.05$ (blue), $\chi/\pi = 0.00$ (orange), $\chi/\pi = 0.05$ (green), $\chi/\pi = 0.10$ (red), $\chi/\pi = 0.15$ (purple),  $\chi/\pi = 0.20$ (brown), and $\chi/\pi = 0.25$ (pink). 
		Black circles show snapshot based estimators for the transient Hall polarization at times $\tau/t_x = 2.0,~2.5,\dots,6.0$, using $N = 10^4$ snapshots. 
		The error bars indicate two standard deviations of the mean value of $P_\mathrm{H}$,  $\pm2 \sigma_{P_\mathrm{H}} / \sqrt{N}$.
		(b) Transient dynamics in the transverse polarization $p_y$, with error bars corresponding to two standard deviations, $\pm2 \sigma_{p_y} / \sqrt{N}$.
		(c) Transient dynamics in the longitudinal current $j_x$.
		The data for different values of $\chi$ are vertically offset by $0$ ($\chi/\pi = -0.05$), $-2$ ($\chi/\pi = 0.00$), $-4$ ($\chi/\pi = 0.05$), $-6$ ($\chi/\pi = 0.10$), $-8$ ($\chi/\pi = 0.15$), $-10$ ($\chi/\pi = 0.20$), and $-12$ ($\chi/\pi = 0.25$) for the purpose of a clear presentation.
		The errors $\pm2 \sigma_{j_x} / \sqrt{N}$ for the longitudinal current are comparable to the size of the symbols (black circles). 
		Hence, they are not explicitly shown in the figure.
		Note that snapshot based estimators of the Hall polarization $P_\mathrm{H} = p_y/j_x$ are computed from the transverse polarization $p_y$ and the longitudinal current $j_x$, which can be directly sampled from the snapshot data, with propagated uncertainties.
		All sets of model parameters considered in this figure correspond to the superfluid Meissner phase. 
	} 
	\label{fig:hall_dynamics}
\end{figure}
In the following, we demonstrate that the Hall polarization $P_\mathrm{H}$ can be effectively sampled from snapshot data.
For this, we employ a suitable and realistic quench protocol~\cite{buser_21}, considering a system with open boundaries and focusing on the Meissner phase.
The protocol starts off with the ground state of the flux-ladder Hamiltonian~\eqref{eq:fluxladderhamiltonian}, which is obtained from a preliminary single-site density-matrix renormalization-group simulation~\cite{white_92, schollwoeck_11}, using subspace expansion~\cite{hubig_15}.
A transient longitudinal current is induced by statically tilting the ladder in the longitudinal direction.
Explicitly, at time $\tau = 0$, the ladder is instantaneously subjected to an additional linear potential $\hat{V}_x = \mu_x \sum_m\sum_r r \hat{n}_{r,m}$. 
Subsequently, the state evolves according to $\hat{H}+\hat{V}_x$.
Actual snapshots are then taken in the transient regime.
Concretely, at times $\tau/t_x = 2.0,~2.5,\dots,~6.0$, the eight most central rungs of the ladder are current-transformed as discussed in the previous section and $N=10^4$ snapshots are sampled from the time-evolved and transformed states.
Note that for the time evolution of matrix-product states, we employ the two-site variant of the time-dependent variational principle algorithm~\cite{haegeman_11, paeckel_19}, as implemented in the SyTen toolkit~\cite{hubigthesis_17, syten}.
%

%
The snapshot results for the Hall polarization are shown in Fig.~\ref{fig:hall_dynamics}.
The considered model parameters, namely the interparticle interaction strength $U/t_x = 2$, the interleg hopping strength $t_y/t_x=1.6$, the particle filling $\nu=0.8$, and the various values of the magnetic flux $\chi/\pi=-0.05,~0.00,\dots,0.25$ all correspond to the Meissner phase.
Moreover, the tilt parameter is chosen to be $\mu_x/t_x=0.1$ and in the numerical simulations, a site-local cutoff at at most three bosons per lattice site is employed.
%

%
Figure~\ref{fig:hall_dynamics}(a) shows the transient Hall polarization $P_\mathrm{H}$ for the various values of the magnetic flux $\chi$ after inducing the tilt dynamics at time $\tau=0$.
The solid lines show exact results that are directly calculated in the central eight rungs from the time-evolved matrix-product states~\cite{buser_21}, for increasing values of the magnetic flux $\chi$ from top to bottom.
For the different values of $\chi$, the transient dynamics in the Hall polarization $P_\mathrm{H}$ reveal clear oscillations with different time averages. 
At times $\tau/t_x = 2.0,~2.5,\dots,~6.0$, black circles show snapshot based data for $P_\mathrm{H}$, which are obtained from $N=10^4$ snapshots, with error bars corresponding to two standard deviations, $\pm2 \sigma_{P_\mathrm{H}} / \sqrt{N}$.
The time-dependent snapshot based data for $P_\mathrm{H}$ are in accordance with the exact results.
%

%
Figure~\ref{fig:hall_dynamics}(b) and Fig.~\ref{fig:hall_dynamics}(c) show the tilt dynamics in the transverse polarization $p_y$ and in the longitudinal current $j_x$, respectively.
The solid lines show exact results, using the color code from Fig.~\ref{fig:hall_dynamics}(a), and black circles indicate the snapshot based mean values. 
In Fig.~\ref{fig:hall_dynamics}(b), the error bars correspond to two standard deviations, $\pm2 \sigma_{p_y} / \sqrt{N}$.
The data corresponding to different values of the magnetic flux $\chi$ shown in Fig.~\ref{fig:hall_dynamics}(b) are set off vertically for the purpose of a clear presentation.
Note that the snapshot based Hall polarization $P_\mathrm{H} = p_y/j_x$ shown in Fig.~\ref{fig:hall_dynamics}(a) is computed from the transverse polarization $p_y$ and the longitudinal current $j_x$ shown in Fig.~\ref{fig:hall_dynamics}(b) and Fig.~\ref{fig:hall_dynamics}(c).
The transverse polarization $p_y$ and the longitudinal current $j_x$ are directly sampled from the snapshots, as discussed in Sec.~\ref{sec:hall_response:snapshot_approach}.
The uncertainties of $P_\mathrm{H}=p_y/j_x$ are propagated accordingly.
\subsection{Estimation of the Hall voltage}
\label{sec:hall_response:hall_voltage}
\begin{figure}
	\centering
	\includegraphics{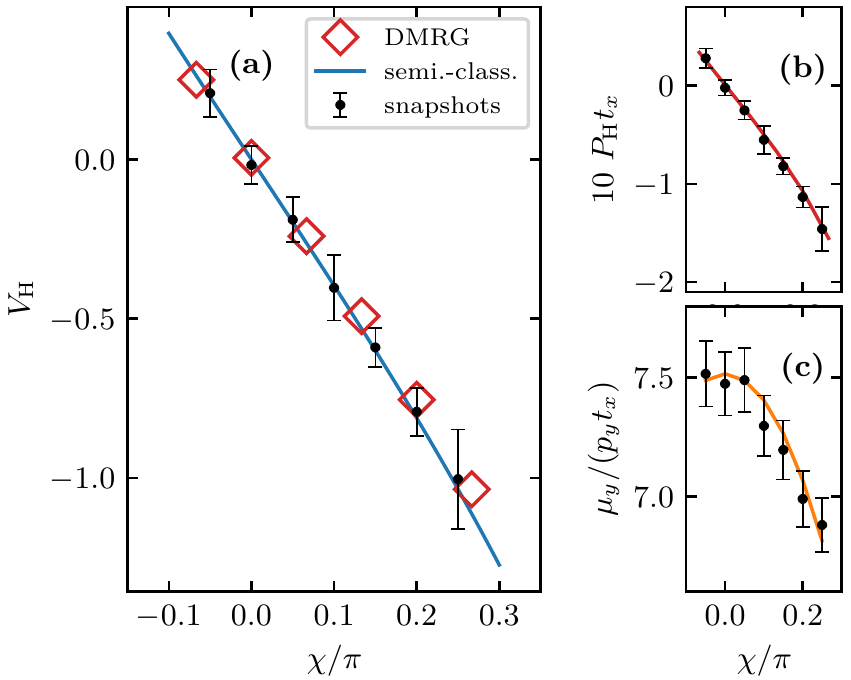}
	\caption{
		Snapshot based estimation of the Hall voltage.
		Considering $U/t_x = 2 $, $t_y/t_x = 1.6$,  $\nu=0.8$, and a local bosonic cutoff at at most three particles per lattice site.
		(a) Hall voltage $V_\mathrm{H}$ versus magnetic flux $\chi$ as computed from ground-state ring-ladder simulations (DMRG) and from a semi-classical coherent-state ansatz (semi.-class.)~\cite{buser_21}.
		Black circles and error bars (corresponding to plus/minus two standard deviations) show the snapshot based estimators for $V_\mathrm{H}$, using the snapshot results for the Hall polarization $P_\mathrm{H}$ shown in Fig.~\ref{fig:hall_dynamics} and additional snapshots results for $\mu_y/p_y$, as discussed in the main text and shown in (c).
		(b) Hall polarization $P_\mathrm{H}$ versus magnetic flux $\chi$. 
		Black dots show time-averaged results from Fig.~\ref{fig:hall_dynamics} and the error bars show the propagated uncertainty corresponding to plus/minus two standard deviations.
		The red solid line shows the result obtained from a ground-state calculations in ring ladders with periodic boundary conditions.  
		(c) Black circles show averaged snapshot results for $\mu_y/p_y$, for $N=10^4$ and $\mu_y/t_x = 1$, with error bars corresponding to two standard deviations, $\pm2 \sigma_{\mu_y/p_y} / \sqrt{N}$. 
		The orange solid line shows results that are directly calculated from the underlying ground states.
		Note that the ring-ladder approch and the semi-classical ansatz are both discussed in detail in Ref.~\cite{buser_21}. 
		All of the model parameters considered in this figure correspond to the Meissner phase. 
	} 
	\label{fig:snap_hall_voltage}
\end{figure}
Finally, we turn to the snapshot based estimation of the Hall voltage $V_\mathrm{H}$ via $V_\mathrm{H}=P_\mathrm{H}(\mu_y/p_y)$.
First, the Hall polarization is calculated by time averaging the snapshot results for $P_\mathrm{H} = p_y/j_x$ in the transient dynamics that are induced by a static longitudinal tilt with $\mu_x/t_x=0.1$.
Explicitly, for each value of the magnetic flux $\chi/\pi=-0.05,~0.00,\dots,0.25$, we obtain $P_\mathrm{H}$ as the time average of all of the snapshot results which are shown in Fig.~\ref{fig:hall_dynamics}, that is, for values of $\tau / t_x = 2.0,~2.5,\dots,~6.0$.
Second, snapshot results for $(\mu_y/p_y)$ are obtained from an additional finite-bias calculation.
For this, the ground state of $\hat{H}+\mu_y \hat{P}_y$ is optimized in a density-matrix renormalization-group simulation and, subsequently, standard Fock-basis snapshots are drawn from the finite-bias ground state.
These snapshots directly give rise to $(\mu_y/p_y)$.
%

%
For the model parameters that are also considered in the previous Sec.~\ref{sec:hall_response:static_tilt}, Fig.~\ref{fig:snap_hall_voltage} presents snapshot results concerning the Hall voltage $V_\mathrm{H}$.
As a function of the magnetic flux $\chi$, Fig.~\ref{fig:snap_hall_voltage}(a) shows $V_\mathrm{H}$ as independently obtained from density-matrix renormalization-group simulations in ring ladders with periodic boundary conditions~(DMRG), from semi-classical calculations~(semi.-class.), and from matrix-product-state based snapshots of systems with open boundaries~(snapshots)~\cite{buser_21}.
Importantly, the snapshot based results for the Hall voltage are in accordance with the ring-ladder simulations and with the semi-classical calculations.
We recap that the snapshot results are computed by means of $V_\mathrm{H}=P_\mathrm{H}(\mu_y/p_y)$.
In Fig.~\ref{fig:snap_hall_voltage}(b) the black circles show snapshot results for $P_\mathrm{H}$ as a function of the magnetic flux $\chi$, which are obtained from the static-tilt protocol discussed in Sec.~\ref{sec:hall_response:static_tilt} with $N=10^4$ snapshots per data point. 
The error bars correspond to two standard deviations. 
The red solid line in Fig.~\ref{fig:snap_hall_voltage}(b) shows ground-state results for $P_\mathrm{H}$ that are directly computed in ring ladders with periodic boundary conditions, employing density-matrix renormalization-group simulations~\cite{hubig_15}.
Using $N=10^4$ snapshots for each value of $\chi$, the black circles in Fig.~\ref{fig:snap_hall_voltage}(c) show finite-bias ground-state results for $\left(p_y/\mu_y\right)$, considering $\mu_y=t_x$ in a ladder with $L=40$ rungs and open boundaries.
The error bars shown in Fig.~\ref{fig:snap_hall_voltage}(c) correspond to two standard deviations, $\pm 2 \sigma_{\mu_y/p_y} / \sqrt{N}$.
The orange solid line shows results for $\left(p_y/\mu_y\right)$ that are directly computed in the underlying ground states.
%

%
Overall, the agreement between snapshot, semi-classical, and ring-ladder results shows that the Hall polarization $P_\mathrm{H}$ and the Hall voltage $V_\mathrm{H}$ can be faithfully sampled in experimentally feasible quench and finite-bias simulations.
It is worth noting that the predicted uncertainties shown in Fig.~\ref{fig:snap_hall_voltage} are based on $N=10^4$ snapshots per data point, which is a feasible number in experiments~\cite{chiu_19}.
\section{Summary}
\label{sec:summary}
In this paper, we employed snapshots that are sampled from matrix-product states in order to study local particle-current statistics and the Hall response of interacting bosons in two-leg flux ladders.
The snapshot-sampling routine, which is based on the approach outlined by Ferris and Vidal in Ref.~\cite{ferris_12}, was discussed in detail. 
Our implementation ensures that the $U(1)$ quantum symmetry corresponding to the particle-number conservation of the flux-ladder Hamiltonian~\eqref{eq:fluxladderhamiltonian} is preserved while snapshots are sampled.
A particular focus was on the exploitation of unitary two-site transformations that enable the sampling of local-particle currents in bosonic lattice models.
%

%
In order to exemplify the sampling approach, we concentrated on characteristic rung-current patterns in the vortex-lattice\textsubscript{1/2} phase of the flux-ladder Hamilitonian~\eqref{eq:fluxladderhamiltonian}.
It was shown that the local rung-current statistics and the probability distribution of the staggered rung-current operator can be faithfully reconstructed from sampled ground-state snapshots.
%

%
Moreover, we computed the Hall polarization and the Hall voltage from theoretical snapshot data.
Concretely, the snapshots were obtained from time-evolved matrix-product states, following feasible quench and finite-bias simulations~\cite{buser_21}.
A suitable pattern of unitary two-site transformations enabled the simultaneous sampling of the particle-current profiles and the particle-density profiles, that are both required for the characterization of the Hall response.
It was shown that the snapshot based estimators for the Hall polarization and for the Hall response are in perfect accordance with the results obtained from density-matrix renormalization-group and semi-classical calculations.
%

%
The theoretical methods employed in this work can guide future experiments with quantum-gas microscopes measuring the Hall response in quasi-one-dimensional flux-lattice systems.
They might prove useful for further explorations of the Hall response in the two-dimensional regime as well~\cite{repellin_20}.
Our sampling approach paves the way for future snapshot based studies of particle currents in interesting quantum states in flux ladders~\cite{petrescu_17, strinati_17, strinati_19}, flux cylinders~\cite{palm_20}, and beyond.
\section{Acknowledgments}
\label{sec:acknowledgments}
We thank Annabelle Bohrdt, Thierry Giamarchi, Sebastian Greschner, Fabian Heidrich-Meisner, Sam Mardazad, Sebastian Paeckel, and Leticia Tarruell for inspiring and helpful discussions.
We acknowledge funding by the Deutsche Forschungsgemeinschaft (DFG, German Research Foundation)
under Germany’s Excellence Strategy -- EXC-2111 -- 390814868 and via Research Unit FOR 2414 under project number 277974659.
\bibliography{snaps_bib}
\end{document}